\documentclass[aps,prb,twocolumn,showpacs,floatfix]{revtex4}
\usepackage{amssymb}
\usepackage{amsmath}
\usepackage{graphicx}

\begin{document}

\title{Magnetic susceptibility of topological nodal semimetals}

\author{G.~P.~Mikitik}
\affiliation{B.~Verkin Institute for Low Temperature Physics \&
Engineering, Ukrainian Academy of Sciences, Kharkov 61103,
Ukraine}

\author{Yu.~V.~Sharlai}
\affiliation{B.~Verkin Institute for Low Temperature Physics \&
Engineering, Ukrainian Academy of Sciences, Kharkov 61103,
Ukraine}

\begin{abstract} Magnetic susceptibility of the topological Weyl, type-II Weyl, Dirac, and line node semimetals is theoretically investigated. Dependences of this susceptibility on the chemical potential, temperature, direction and magnitude of the magnetic field are found. The obtained results show that magnetic measurements can be very useful in investigating these semimetals.
As an example, we calculate magnetic susceptibility of Cd$_3$As$_2$, Na$_3$Bi, and Ca$_3$P$_2$.
\end{abstract}


\maketitle

\section{Introduction}

In nodal semimetals the conduction and valence bands of electrons touch at points or along lines in three-dimensional quasi-momentum space. \cite{wan,balents,young} There are several types of these nodal materials. They are the Weyl, Dirac and line node semimetals. In the Weyl semimetals, the electron bands contact at discrete (Weyl) points of the Brillouin zone and disperse linearly in all directions around these critical points. The same type of the band contact occurs in the Dirac semimetals, but the bands are double degenerate in spin, i.e., a Dirac point can be considered as a couple of the Weyl points overlaping in the quasi-momentum space. In the line node semimetals the conduction and valence bands touch along lines in the Brillouin zone and disperse linearly in directions perpendicular to these lines. A number of various the Weyl, Dirac, and line node semimetal were predicted and discovered experimentally in recent years. \cite{wang,neupane,borisenko,jeon,liang,ali,liu,z.wang,liu1,h.weng,
huang,shek,hei1,pie,weng,xie,kim,yu,yama,phil,schoop,bian,chen} In this paper we call attention to the fact that the magnetic susceptibility of electrons in topological nodal semimetals exhibits a giant anomaly that can be useful in experimental investigations of these material.

As early as 1989,\cite{m-sv} all the types of electron-band degeneracy in three-dimensional crystals were found that lead to a giant anomaly in their magnetic susceptibility $\chi$ in weak magnetic fields $H$. This anomaly occurs in the orbital part of the susceptibility and is characterized by divergence of $\chi$ at low temperatures $T$ when the chemical potential $\zeta$ of the electrons approaches the band-degeneracy energy $\varepsilon_d$. The anomaly is due to virtual interband transitions of electrons under the action of the magnetic field. These transitions give a contribution to the susceptibility which is inversely proportional to a gap between the electron bands, and so this contribution is large for nearly degenerate electron states. The type of the divergence of $\chi$ is determined by the character of band-degeneracy lifting in the vicinity of $\varepsilon_d$. Below we consider only those ``$\chi$-divergence'' types of the electron-band degeneracy that are appropriate to the topological semimetals.

The first type of the degeneracy is just a Dirac or Weyl point of the electron energy spectrum in a three-dimensional crystal. In this case the susceptibility at low temperatures and weak magnetic fields diverges logarithmically, $\chi\propto \ln|\zeta-\varepsilon_d|$. \cite{m-sv} Later, the same divergence of $\chi$  was also obtained by Kashino and Ando for the case of an isotropic Dirac point. \cite{kosh} The giant anomaly in $\chi$ for strong magnetic fields  was studied in Ref.~\onlinecite{m-sh} (see also Ref.~\onlinecite{roy}), and a dependence of $\chi$ on $H$ was found for such fields. In this paper, we present the results of Refs.~\onlinecite{m-sv} and \onlinecite{m-sh} that are appropriate to  Dirac and Weyl semimetals including the so-called \cite{sol} type-II Weyl semimetals, and based on these results, we analyze the magnetic susceptibility (and magnetic torque) of these materials in detail.

The second type of the band degeneracy leading to the giant anomaly in $\chi$ is the band-contact lines \cite{herring} in the Brillouin zone of crystals. In this case the degeneracy energy $\varepsilon_d$ changes along the line and reaches its maximum $\varepsilon_{max}$ and minimum $\varepsilon_{min}$ values at certain points ${\bf p}_{i}$ of the line. The giant anomaly of $\chi$ is determined by the electron states located near these points ${\bf p}_{i}$ and occurs at low temperatures and weak magnetic fields only when $\zeta$ approaches $\varepsilon_{max}$ from below or $\varepsilon_{min}$ from above,\cite{m-sv} $\chi\propto (\zeta-\varepsilon_{min})^{-1/2}$ or $\chi\propto (\varepsilon_{max}-\zeta)^{-1/2}$. At these critical energies $\varepsilon_{min}$ and $\varepsilon_{max}$, the so called electron topological transitions of $3\frac{1}{2}$ kind occur in metals. \cite{m-sh14} For  strong magnetic fields, this type of the giant anomaly in $\chi$ was analyzed in Ref.~\onlinecite{m-sh}, and a $H$-dependence of the magnetic susceptibility was derived for this case. However, it is necessary to emphasize that these results for $\chi$ in weak and strong magnetic fields were obtained  under the implicit assumption that the temperature $T$ and the characteristic spacing between the Landau subbands in the magnetic field, $\Delta\varepsilon_H$, are much less than the difference $\varepsilon_{max}-\varepsilon_{min}\equiv 2\Delta$. Then, contributions of the different critical points ${\bf p}_i$ in the band-contact line to the magnetic susceptibility can be considered independently. However, in the line node semimetals, the difference $2\Delta$ is assumed to be small, and so the results  of Refs.~\onlinecite{m-sv} and \onlinecite{m-sh} for the case of band-contact lines are applicable to the semimetals only at not-too-high  $T$ and $H$.

Recently, \cite{kosh15} the magnetic susceptibility $\chi$ of electrons in weak magnetic fields was estimated for a semimetal with a band-contact ring lying at the constant energy $\varepsilon_d$ (i.e. $\Delta=0$ for the ring). This ring was described by the particular model proposed in Ref.~\onlinecite{balents}, and it was found that $\chi(\zeta)$ is proportional to the delta function $\delta(\zeta- \varepsilon_d)$.\cite{kosh15} In this paper, we admit any values of $\Delta$, and complementing the results of Refs.~\onlinecite{m-sv} and \onlinecite{m-sh}, we calculate the magnetic susceptibility (and magnetic torque) of the line node semimetals with the band-contact lines of arbitrary shapes in the weak and strong magnetic fields for the cases when $\Delta$ is less than $T$ or $\Delta\varepsilon_H$.

The paper is organized as follows: In Sec.~\ref{weyl}, we describe dependences of the magnetic susceptibility of electrons in the Weyl and Dirac semimetals on the chemical potential, temperature, direction and magnitude of the magnetic field. As an example, we analyze the magnetic susceptibility of electrons in Cd$_3$As$_2$ and Na$_3$Bi. In Sec.~\ref{line}, the magnetic susceptibility in the line node semimetals is studied at any interrelations between $T$, $\Delta\varepsilon_H$, and $\Delta$. We also apply the obtained results to Ca$_3$P$_2$. A brief summary of our results is presented in Sec.~\ref{conc}.

\section{Weyl and Dirac semimetals}\label{weyl}

It has been shown recently that Cd$_3$As$_2$ \cite{wang,neupane,borisenko,jeon,liang,ali,liu} and Na$_3$Bi \cite{z.wang,liu1} fall into the class of the topological Dirac semimetals, while Weyl semimetal phase is realized in noncentrosymmetric crystals TX \cite{h.weng,
huang,shek} where T is Nb or Ta and X is As or P. We start the theoretical study of the magnetic susceptibility for such topological semimetals with a description of the electron spectrum near the Dirac and Weyl points.

\subsection{Spectrum}\label{spectr}

The most general ${\bf k}\cdot{\bf p}$
Hamiltonian $\hat H$ for the conduction and valence electrons in the vicinity of a Dirac point  has the form:\cite{m-sv,m-sh}
 \begin{eqnarray}\label{1}
\hat H=\left (\begin{array}{cccc} E_{c} & R & 0 & S \\ R^* & E_{v} & -S & 0 \\ 0 & -S^* & E_{c} & R^* \\ S^* & 0 & R & E_{v} \\
\end{array} \right),
 \end{eqnarray}
where
 \begin{eqnarray}\label{2}
E_{c,v}&=&\varepsilon_d+{\bf v}_{c,v}\cdot {\bf p}, \nonumber \\
R&=&{\bf r}\cdot {\bf p}, \\
S&=&{\bf s}\cdot {\bf p}.  \nonumber
  \end{eqnarray}
Here the quasi-momentum ${\bf p}$ is measured from the Dirac point;  ${\bf v}_{c,v}$ are intraband and ${\bf r}$ and ${\bf s}$ are interband matrix elements of the velocity operator calculated at ${\bf p}=0$; the vectors ${\bf v}_{c,v}$ are real, while ${\bf r}$ and ${\bf s}$ are generally complex quantities. Hamiltonian (\ref{1}) takes into account a twofold spin degeneracy of electron bands in centrosymmetric crystals. If one set $S=0$, this Hamiltonian (e.g., its upper $2\times 2$ block)  describes the electron states near the  Weyl points.

Diagonalization of the  Hamiltonian (\ref{1}), (\ref{2}) gives  the   dispersion  relations for  the  electron bands in the   vicinity of  the Dirac (Weyl) point:
 \begin{eqnarray}\label{3}
 \varepsilon_{c,v}&=&\varepsilon_d+{\bf a}\cdot{\bf p}+E_{c,v}, \\
 E_{c,v}&=&\pm\{({\bf a'}{\bf k})^2+|R|^2+|S|^2 \}^{1/2}, \nonumber
 \end{eqnarray}
where  the   following notations have been introduced:
 \begin{eqnarray*}
 {\bf a}=({\bf v}_c+{\bf v}_v)/2;\ \ \ {\bf a'}=({\bf v}_c-{\bf v}_v)/2.
 \end{eqnarray*}
Equation (\ref{3}) shows that $E_{c,v}^2$ is a quadratic form in the components of  vector ${\bf p}$. Hereafter we choose the coordinate axes along principal directions of  this form. Let  $b_{ii}$ ($i=1, 2, 3$)  be its principal values, i.e.,
\[E_{c,v}^2=b_{11}p_1^2+b_{22}p_2^2+b_{33}p_3^2,\]
where $b_{ii}$ are expressible in terms of  the   components of  the   vectors  ${\bf a'}$, ${\bf r}$,  ${\bf s}$. The scaling of coordinate axes, $\tilde p_i=p_i\sqrt{b_{ii}}$, transforms Eq.~(\ref{3}) into the form that depends only on the constant dimensionless vector $\tilde {\bf a}$:
 \begin{eqnarray}\label{4}
 \varepsilon_{c,v}&=&\varepsilon_d+\tilde{\bf a}\cdot\tilde{\bf p}\pm |\tilde {\bf p}|,
 \end{eqnarray}
where its components are defined by $\tilde a_i=a_i/\sqrt{b_{ii}}$. When  the length of $\tilde {\bf a}$ is less than unity,
\begin{eqnarray}\label{5}
\tilde a^2=\frac{a_1^2}{b_{11}}+ \frac{a_2^2}{b_{22}} +\frac{a_3^2}{b_{33}} < 1,
 \end{eqnarray}
the dispersion relations $\varepsilon_{c,v}({\bf p})$ looks like in Fig.~1a, and the Fermi surface at $\zeta=\varepsilon_d$ is a point. When $\tilde a^2>1$, there is always a direction in ${\bf p}$-space along which the dispersion relations $\varepsilon_{c,v}({\bf p})$ looks like in Fig.~1b, and at $\zeta=\varepsilon_d$ the open electron and hole pockets of the Fermi surface touch. Thus, the case $\tilde a^2>1$ corresponds to the type-II Weyl (or Dirac) points. \cite{sol} We shall see below that the magnetic susceptibilities at $\tilde a^2>1$ and $\tilde a^2<1$ are essentially different.

\begin{figure}[tbp] % %%%%%%%%%%%%%%%%%%%%%%%%%%%%%%%%%%%%%
 \centering  \vspace{+9 pt}
\includegraphics[scale=.90]{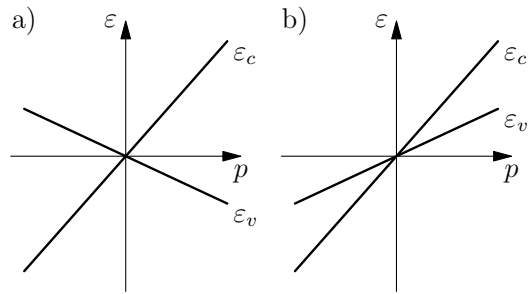}
\caption{\label{fig1} Dispersion relations $\varepsilon_c(p)$ and $\varepsilon_v(p)$ of the electron energy bands in the vicinity of a degeneracy point in the cases of $\tilde a^2<1$ (a) and $\tilde a^2>1$ (b), Eq.~(\ref{4}).
 } \end{figure}   %%%%%%%%%%%%%%%%%%%%%%%%%%%%%%%%%%%%%%%%%%

In the magnetic field ${\bf H}={\bf n}H$ directed along the unit vector ${\bf n}$, the spectrum of electrons described by Hamiltonian (\ref{1}), (\ref{2}) has the form:\cite{m-sh} \begin{eqnarray}\label{6}
 \varepsilon_{c,v}^l(p_n)&=&\varepsilon_d +vp_n \pm \left[\frac{e\hbar \alpha H}{c}\,l+L\cdot (p_n)^2 \right]^{1/2},
 \end{eqnarray}
where $e$ is the absolute value of the electron charge;  $l=0,1,2,\dots$; $p_n= {\bf p}{\bf n}$ is the component of the quasi-momentum along the magnetic field,
\begin{eqnarray}\label{7}
\alpha&=&\frac{2R_n^{3/2}}{b_{11}b_{22}b_{33}\tilde{\bf n}^2}, \nonumber \\
L&=&\frac{R_n}{b_{11}b_{22}b_{33}\tilde{\bf n}^4}, \nonumber \\
R_n\!\!&=&\sum_{i,j=1}^{3}\kappa^{ij}n_in_j, \\
\kappa^{ij}\!\!&=&\frac{b_{11}b_{22}b_{33}}{(b_{ii}b_{jj})^{1/2}} \left[(1-\tilde a^2)\delta_{ij}+\tilde a_i\tilde a_j\right],\nonumber \\
v&=&\frac{(\tilde {\bf a}\tilde{\bf n})}{\tilde{\bf n}^2}, \nonumber
 \end{eqnarray}
and the components of the vector $\tilde{\bf n}$ are determined by the relation: $\tilde n_i=n_i/\sqrt{b_{ii}}$. For given ${\bf n}$, the spectrum (\ref{6}) describing the Landau subbands $\varepsilon^l_{c,v}(p_n)$ exists if $R_n>0$. When $\tilde a^2<1$, the $R_n$ is positive at {\it any} direction of the magnetic field.
We also emphasize that in most publications dealing with the electron energy spectrum in a magnetic field for Dirac (Weyl) points, the simplest (isotropic) case, in which ${\bf a}=\tilde{\bf a}=0$, is usually considered. However, this case can occur only if the Dirac or Weyl point coincides with a highly symmetric point  of the Brillouin zone. Generally the vector ${\bf a}$ always differs from zero, see, e.g., Sec.~\ref{IIc}.

Interestingly, formula (\ref{6}) is equivalent to the equation
\begin{equation}\label{8}
S(\varepsilon^{l},p_n)=\frac{2\pi\hbar e H}{c}l,
\end{equation}
where $S(\varepsilon^{l},p_n)$ is the area of the cross-section of the constant-energy surface $\varepsilon_{c,v}({\bf p})=\varepsilon^l$ by the plane $p_n=$ const. For the Dirac points,  this expression  looks like the well-known semiclassical quantization condition \cite{Sh}
\begin{equation}\label{9}
S(\varepsilon^{l},p_n)=\frac{2\pi\hbar e H}{c}\left(l+\frac{1}{2}\pm \frac{gm_*}{4m}\right)
\end{equation}
if one takes $g=2m/m_*$. Here $g$ is the electron $g$ factor, while $m$ and $m_*=(1/2\pi)\partial S(\varepsilon,p_n)/\partial\varepsilon$ are the electron and cyclotron masses, respectively. For the two-bands Hamiltonian (\ref{1}), (\ref{2}), the orbital  $g$ factor is indeed equal to $2m/m_*$. \cite{g1,g2,g3} This means that this large $g$ factor has been implicitly taken into account in deriving  Eq.~(\ref{6}). However, we have neglected the direct interaction $(e\hbar/mc){\bf s}{\bf H}$ of the electron spin ${\bf s}$ with the magnetic field ${\bf H}$ and a contribution of the bands different from ``c'' and ``v'' to the electron $g$ factor. The impact of all these effects on the Landau subbands and on the magnetic susceptibility is relatively small and is of the order of $m_*/m\sim |\zeta- \varepsilon_d|/mV^2$ where $\zeta$ is the chemical potential, and $V$ is the velocity determining the characteristic slope of Dirac cone, $V\sim |d\varepsilon_{c,v}/dp| \sim \sqrt{b_{ii}}$. With these effects, a small splitting of the Landau subbands appears for the  Dirac point. Hence, if a noticeable splitting of the subbands occurs in an experiment, this is a signal that one should take into account more than two bands in the Hamiltonian describing the Dirac points. In other words, terms of higher orders in $p_i$ than the linear ones should be incorporated in Hamiltonian (\ref{1}), (\ref{2}) in this case.

\subsection{Magnetic susceptibility}\label{IIb}

We define the magnetic susceptibility tensor as $-\partial^2 \Omega/\partial H_i \partial H_j$ where $\Omega$ is the $\Omega$ potential per unit volume of a crystal. When the chemical potential $\zeta$ lies near $\varepsilon_d$, the total magnetic susceptibility tensor $\chi^{ij}_{\rm tot}$ consists of its special part $\chi^{ij}$ determined by the electron states located near the Dirac (Weyl) point and a practically {\it constant} background term $\chi^{ij}_0$ specified by electron states located far away from this point,
\[
\chi_{\rm tot}^{ij}=\chi^{ij} +\chi^{ij}_0.
\]
It is the special part of the tensor that is responsible for dependences of the susceptibility on the chemical potential, temperature, and magnitude of the magnetic field.

In weak magnetic fields $H\ll H_T$, when the characteristic spacing  $\Delta\varepsilon_H$ between the Landau subbands is much less than the temperature $T$, the  susceptibility $\chi^{ij}$ is practically  independent of $H$. On the other hand, at $H>H_T$, when $\Delta\varepsilon_H>T$, a noticeable $H$-dependence of $\chi^{ij}$ appears, and it is more convenient to consider the magnetization $M$ rather than the magnetic susceptibility in this case. The background term $\chi^{ij}_0$ remains constant at all magnetic fields.

According to Eqs.~(\ref{6}) and (\ref{7}), we have the following estimate for the spacing $\Delta\varepsilon_H$ between the Landau subbands of electrons in the magnetic field: $\Delta\varepsilon_H\sim (e\hbar HV^2/c)^{1/2}$, and hence
\begin{equation}\label{10}
H_T\sim \frac{cT^2}{e\hbar V^2}\approx 0.7\cdot10^{12}\frac{T^2}{V^2},
\end{equation}
where, in the last equality, the velocity $V$ characterizing the slope of the Dirac cone is measured in m/s, the temperature $T$ in K, and $H_T$ in Oe. If $V\sim 10^6-10^5$ m/s, we obtain $H_T\sim 10-1000$ Oe  at $T=4$ K. In other words, at low temperatures, a noticeable dependence of $\chi^{ij}$ on $H$ is expected to occur at sufficiently low magnetic fields in the vicinity of the Dirac and Weyl points.

The special part $\chi^{ij}$ of the magnetic susceptibility associated with a Dirac or Weyl point can be calculated at arbitrary magnetic fields with Eqs.~(\ref{6}) and (\ref{7}).\cite{m-sh} Below
we discuss only the case of the Dirac point and present formulas for $\chi^{ij}$ in weak magnetic fields and for the magnetization $M_i=-\partial \Omega/\partial H_i$ in strong magnetic fields. For the case of the Weyl point, these formulas should be divided by two.

\subsubsection{Weak magnetic fields}

In the region of weak magnetic fields ($H\ll H_T$) and {\it under condition} (\ref{5}), $\tilde a^2<1$,  we find the following  expression for $\chi^{ij}$ per unit volume:\cite{m-sv}
\begin{eqnarray}\label{11}
 \chi^{ij}\!\!=\!-\frac{1}{6\pi^2\hbar}\!\left(\frac{e}{c}\right)^2\!\!\! \frac{\kappa^{ij}}{(b_{11}b_{22}b_{33})^{1/2}}\!\!\!\int_{0}^{\varepsilon_0} \!\!\frac{d\varepsilon}{\varepsilon}[f(-\varepsilon)\!-\!\!f(\varepsilon)],
 \end{eqnarray}
where  $\kappa^{ij}$ is given by Eq.~(\ref{7}), $f(\varepsilon)$ is the Fermi function with the chemical potential $\zeta$,
\begin{eqnarray}\label{12}
  f(\varepsilon)=\left[1+\exp\!\left(\frac{\varepsilon+\varepsilon_d -\zeta}{T}\right)
  \right ]^{-1},
 \end{eqnarray}
and $\varepsilon_0$ is a sufficiently high energy specifying the interval ($\varepsilon_d-\varepsilon_0, \varepsilon_d+\varepsilon_0$)  in  which Eqs.~(\ref{1})--(\ref{3}) are valid. Different choices of $\varepsilon_0$ would change only the background term $\chi^{ij}_0$ which is unimportant and is not calculated here.

Calculating the integral in Eq.~(\ref{11}) in the limit $T\to 0$, we arrive at
\begin{eqnarray}\label{13}
 \chi^{ij}\!\!=\!-\frac{1}{6\pi^2\hbar}\!\left(\frac{e}{c}\right)^2\!\!\! \frac{\kappa^{ij}}{(b_{11}b_{22}b_{33})^{1/2}}\ln\left( \frac{\varepsilon_0}{|\zeta-\varepsilon_d|}\right)\!.
 \end{eqnarray}
Thus, at $\tilde a^2< 1$ the giant anomaly of the magnetic susceptibility is of the logarithmic character in $\zeta-\varepsilon_d$. If $\tilde a^2> 1$, the appropriate $\chi^{ij}$ proves to be a constant independent of $\zeta$, \cite{m-sv} and hence there is no giant anomaly in the magnetic susceptibility for the cases of the type-II Dirac or Weyl semimetals. In the particular case when $\tilde{\bf a}=0$ and  $b_{11}=b_{22}=b_{33}=V^2$, expression (\ref{13}) coincides with the last formula in Eq.~(38) of Ref.~\onlinecite{kosh15}.

Formula (\ref{13}) gives $\chi^{ij}(\zeta,T)$ at $|\zeta- \varepsilon_d|\gg T$. In this region of $\zeta$ the susceptibility is practically independent of $T$. However, at $|\zeta-\varepsilon_d|\lesssim T$ a $T$-dependence of the susceptibility appears. According to Eq.~(\ref{11}), in this region the divergence of $\chi^{ij}(\zeta)$ in Eq.~(\ref{13}) is cut off, and at $\zeta=\varepsilon_d$ we have
\begin{eqnarray}\label{14}
 \chi^{ij}(\varepsilon_d,T)\!=\!-\frac{1}{6\pi^2\hbar}\!\left( \frac{e}{c}\right)^2\!\!\! \frac{\kappa^{ij}}{(b_{11}b_{22}b_{33})^{1/2}}\ln\left( \frac{\varepsilon_0}{Tq}\right)\!,
 \end{eqnarray}
where
\[\ln q\!=\!\!\int_0^{\infty}\!\!\!\!\!\frac{\ln x}{(1+\cosh x)}\, dx=\!\ln(\pi/2)-C_{EM}\!\approx -0.1256,\]
$C_{EM}\approx 0.5772$ is the Euler-Mascheroni constant, and hence $q\approx 0.882$.

\subsubsection{Strong magnetic fields}

\begin{figure}[tbp] % %%%%%%%%%%%%%%%%%%%%%%%%%%%%%%%%%%%%%
 \centering  \vspace{+9 pt}
\includegraphics[scale=.99]{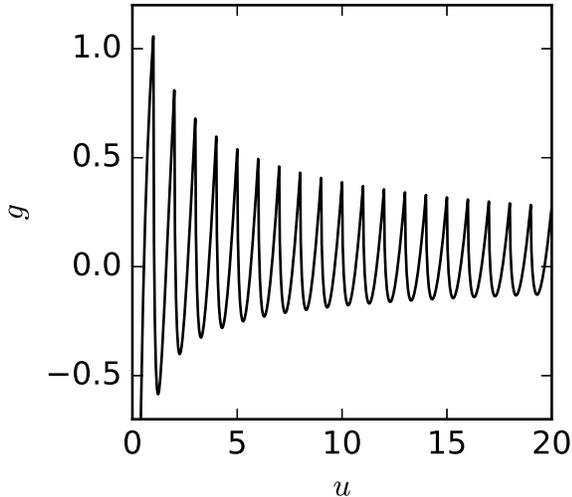}
\caption{\label{fig2} The function $g(u)$, Eq.~(\ref{20}), that describes the oscillations of the magnetization.
 } \end{figure}

In strong magnetic fields, $H\gg H_T$, and at $\zeta=\varepsilon_d$, we find the following expression  for the components $M_i$ of the special part of the magnetization: \cite{m-sh}
\begin{eqnarray}\label{15}
 M_i(\varepsilon_d,\!H)\!\!\!&=&\!\!\!-\frac{e^2 Q_iH}{6\pi^2\hbar\, c^2(b_{11}b_{22}b_{33})^{1/2}}
 \Bigg[A-\frac{1}{4}
  \nonumber \\
 &+&\!\!
 \frac{1}{2}\!\ln\!\!\left(\frac{2\varepsilon_0^2 c}{e\hbar(1-\tilde a^2)R_n^{1/2}H}\!\right)\!
 \Bigg],
 \end{eqnarray}
where $Q_i=\sum_j\kappa^{ij} n_j$, $A\approx 1.50$, $\kappa^{ij}$ and  $R_n$ are given by Eq.~(\ref{7}), and ${\bf n}$ is the unit vector along the magnetic field, $H_j=n_jH$. If $\zeta$ does not coincide exactly with $\varepsilon_d$, the magnetization $M_i$ is the sum of $M_i\!(\varepsilon_d,\!H)$  described by Eq.~(\ref{15}) and the additional term
\begin{eqnarray}\label{16}
\delta M_i&=&\frac{e^2}{2\pi^2c^2\hbar} \frac{Q_iH}{(b_{11}b_{22} b_{33})^{1/2}}\Bigg\{u+2\sum_{m=1}^M\Bigg[\sqrt{u(u-m)} \nonumber \\
&-&2m\ln\left(\frac{\sqrt{u}+\sqrt{u-m}}{\sqrt{m}} \right)\Bigg]\Bigg\},
 \end{eqnarray}
where
\begin{equation}\label{17}
 u\equiv  \frac{(\zeta-\varepsilon_d)^2 c}{2e\hbar(1-\tilde a^2)R_n^{1/2}H}=\frac{S_{\rm ex}c}{2\pi\hbar e H},
 \end{equation}
$M\equiv [u]$ is the integer part of $u$, and $S_{\rm ex}$ is the area of the extremal cross-section of the constant-energy surface $\varepsilon_{c,v}({\bf p})=\zeta$ by the plane $p_n=$ const. which is perpendicular to the magnetic field. The extremum is found relative to $p_n$. The integer $M$ is the number of the Landau subbands occupied by electrons in the conduction band or by holes in the valence band. At $u<1$ the term $\delta M_i$ reduces to the constant
\begin{equation}\label{18}
\delta M_i=\frac{e}{4\pi^2c\hbar^2} \frac{Q_i(\zeta-\varepsilon_d)^2}{(b_{11}b_{22} b_{33})^{1/2}(1-\tilde a^2)R_n^{1/2}}.
\end{equation}
When $u$ increases and $u>1$, oscillations of the magnetic moment   appear. According to formulas (\ref{13}), (\ref{15}), and (\ref{16}), the magnetization $M_i$ can be represented as follows:
\begin{eqnarray}\label{19}
M_i\!=\!\chi^{ij}_{H\to 0}\!\!\cdot\!n_jH\!\!+\!\frac{e Q_i(\zeta- \varepsilon_d)^2 g(u)}{12\pi^2\!\hbar^2 c(b_{11}b_{22}b_{33})^{\!1\!/2}\!(1-\tilde a^2)R_n^{1/2}},~~
 \end{eqnarray}
where $\chi^{ij}_{H\to 0}$ is the susceptibility in the weak field region, Eq.~(\ref{13}), and $g(u)$ is the function which is independent of the parameters of the Dirac point:
\begin{eqnarray}\label{20}
g(u)=-\frac{\ln(2\sqrt u) +A-\frac{1}{4}}{u}+3+
  \frac{6}{u}\sum_{m=1}^M\Bigg[\sqrt{u(u-m)} \nonumber \\
-2m\ln\left(\frac{\sqrt{u}+\sqrt{u-m}}{\sqrt{m}} \right)\Bigg].~~~~
 \end{eqnarray}
This function for not-too-high $u$ is shown in Fig.~\ref{fig2}. At $u\gg 1$, the second term in formula (\ref{19}) describes the well-known oscillations in the de Haas- van Alphen effect.\cite{Sh} However, a  phase of the oscillations is shifted by $\pi$ as compared to the usual case. In other words, if one plots $1/H_l$ versus $l$ where $H_l$ is the magnetic field corresponding to $l$-th peak in the magnetization and draws a straight line through these point, this  line  extrapolated to the origin of the coordinates passes through the point $l=0$ rather than through $l=1/2$. This is the distinguishing property of a Dirac (Weyl) point.

   %%%%%%%%%%%%%%%%%%%%%%%%%%%%%%%%%%%%%%%%%%

Consider now the longitudinal magnetization $M_{\parallel}=\sum_{i} M_in_i$ in more detail. In an appropriate formula for this quantity, the coefficient before the oscillating function $g(u)$ can be expressed in terms of the extremal cross-section area $S_{\rm ex}$ of the Fermi surface and of the cyclotron mass $m_*$ corresponding to this cross section. This representation enables one to describe the oscillations of the magnetization even in the case when the chemical potential $\zeta$ is not close to $\varepsilon_d$. As was mentioned in Sec.~\ref{spectr}, in this situation, terms of higher orders in the quasi-momentum than the linear ones should be incorporated in Hamiltonian (\ref{1}), (\ref{2}). This leads to a change of the electron spectrum (\ref{3}) and to a small splitting of the Landau subbands. In this case, the longitudinal magnetization $M_{\parallel}$ can be described by the following formula generalizing Eq.~(\ref{19}):
\begin{eqnarray}\label{21}
M_{\parallel}\!=\!\chi_{\parallel}(H\!\!\to\! 0)\!\cdot\! H\!\!+\!C\cdot\frac{g(u_+)+ g(u_-)}{2} ,
 \end{eqnarray}
where $\chi_{\parallel}(H\!\!\to\! 0)=\sum_{i,j} \chi^{ij}_{H\to 0}n_in_j$ is the longitudinal magnetic susceptibility in weak magnetic fields,
\begin{eqnarray}\label{22}
C&=&\frac{e S_{\rm ex}^{3/2}}{6\sqrt 2\pi^3\!\hbar^2 c|m_*||S''|^{1/2}}, \\
u_{\pm}\!&=&u \pm \frac{\Delta g m_*}{4m}, \nonumber
 \end{eqnarray}
$S''=\partial^2 S(\zeta,p_n)/\partial p_n^2|_{p_{n,\rm ex}}$ is the second derivative of the cross-section area calculated at $p_n$ corresponding to the extremal cross section, and $\Delta g=g-(2m/m_*)$ is the deviation of the $g$ factor from the value $(2m/m_*)$. It is this $\Delta g$ that leads to the splitting of the Landau subbands; the value of $\Delta g$ can be found with a perturbation theory. \cite{g3} As to $S_{\rm ex}$, $m_*$, and $S''$, these quantities can be calculated if corrections to the spectrum (\ref{3}) are known.

It follows from formulas (\ref{13})-(\ref{20}) that the angular dependence of the special part of the magnetization is completely determined by the tensor $\kappa^{ij}$. The components $\chi_0^{ij}$ of the background term are constant, but they generally differ from each other, and so $\chi_0^{ij}$ also give a contribution to the angular dependence of the total magnetization. However, a comparison of the total magnetization or magnetic torque in weak ($H\ll H_T$) and strong ($H\gg H_T$) magnetic fields permits one to eliminate the effect of the background susceptibility $\chi_0^{ij}$ and of the cut-off parameter $\varepsilon_0$ on these quantities. For example, the difference $M_{\parallel,tot}(H)-\chi_{\parallel,tot}H$ is independent of $\chi_0^{ij}$ and $\varepsilon_0$ if the total longitudinal magnetization $M_{\parallel,tot}(H)=\sum_iM_{i,tot}\cdot n_i$ is measured in strong magnetic fields and the total longitudinal susceptibility $\chi_{\parallel,tot}=\sum_{i,j}\chi^{ij}_{tot}n_in_j$ in weak fields. This difference is completely determined by the parameters of the Dirac point, see Sec.~\ref{IIc}.

Finally, let us briefly discuss the effect of impurities in crystals on the magnetization of the electrons with the Dirac spectrum. This effect for a two-dimensional metal was considered in a number of papers, see, e.g., Refs.~\onlinecite{sharapov} and \onlinecite{kosh07}. In the simplest approximation, in which the width $\Gamma$ of the Landau-level broadening associated with electron scattering from impurities is independent of $H$ and is the same for all the levels, the magnetization $M_{im}$ in the sample with the impurities can be obtained from the magnetization $M$ at $\Gamma=0$ as follows:\cite{sharapov}
 \begin{equation}\label{23}
 M_{im}(\zeta)=\int_{\!-\infty}^{\infty}\!\! d\varepsilon M(\varepsilon)P(\varepsilon-\zeta),
 \end{equation}
where $P(\varepsilon)$ is the probability distribution function describing the broadening of a Landau level,
 \begin{equation}\label{24}
P(\varepsilon)=\frac{1}{\pi}\frac{\Gamma}{\varepsilon^2+\Gamma^2}. \end{equation}
Interestingly, if in Eq.~(\ref{23}) the quantities  $M_{im}(\zeta)$ and $M(\varepsilon)$ are replaced by $M(\zeta,T)$ and  $M(\varepsilon,0)$, respectively, and instead of Eq.~(\ref{24}), one takes
 \[
 P(\varepsilon)=\frac{1}{4T}\left[\cosh\left(\frac{\varepsilon}{2T} \right)\right]^{-2}, 
 \]
the obtained formula describes the effect of temperature on the magnetization, see also Sec.~\ref{IIIb}. It is clear from the similarity of the formulas that in the approach based on Eqs.~(\ref{23}) and (\ref{24}), any singularity in the magnetization is smeared by the  impurities. As shown in Ref.~\onlinecite{sharapov}, this approach is equivalent to the introduction of the Dingle temperature $T_D=\pi \Gamma$ in describing the oscillation part of the magnetization.

\subsection{Example: Cd$_3$As$_2$ and Na$_3$Bi}\label{IIc}

As an example, we apply the formulas of Sec.~\ref{IIb} to Cd$_3$As$_2$ and Na$_3$Bi. In these crystals the electron spectra in the vicinities of their Dirac points coincide, \cite{wang,z.wang} and for simplicity, we discuss only Cd$_3$As$_2$ below.

According to Ref.~\onlinecite{wang}, in Cd$_3$As$_2$ in the axis Z-$\Gamma$-Z of its Brillouin zone the two Dirac points with coordinates $p_x=0$, $p_y=0$, $p_z=\pm p_z^c$ exist, and in the vicinity of any of these points the electron spectrum has the form:\cite{wang}
 \begin{eqnarray}\label{25}
 \varepsilon_{c,v}\!\!=\!\varepsilon_d\!+\!a_z\delta p_z\!\pm\!\{(A\delta p_x)^2\!\!+\!(A\delta p_y)^2\!\!+\!b_{zz}(\delta p_z)^2\}^{\!\!1/2}\!\!\!,
 \end{eqnarray}
where $\delta{\bf p}$ is the quasi-momentum measured from the Dirac point, and the constants $p_z^c$, $a_z$, and $b_{zz}$ in the notations of Ref.~\onlinecite{wang} are: $p_z^c=\hbar\sqrt{M_0/M_1}$, $a_z=\pm 2C_1p_z^c/\hbar^2$,  $b_{zz}=(2M_1p_z^c/\hbar^2)^2$. In Cd$_3$As$_2$ the axes 1, 2, and 3 coincide with the coordinate axes x, y, and z. Equation (\ref{25}) shows that the vector ${\bf a}$ is directed along the $z$ axis, and $b_{11}=b_{22}=A^2$.  Then, according to Eq.~(\ref{7}), the tensor $\kappa^{ij}$ is diagonal and has the following nonzero components: $\kappa^{11}=\kappa^{22}\equiv \kappa=A^2(b_{zz}-a_z^2)$, $\kappa^{33}=A^4$, whereas $R_n=\kappa^{33}\cos^2\theta+\kappa \sin^2\theta$ where $\theta$ is the angle between the magnetic field and the $z$ axis. The background-susceptibility tensor $\chi_0^{ij}$ is also diagonal and is defined by the two constants: $\chi_0^{33}$ and $\chi_0^{11}= \chi_0^{22}\equiv \chi_0^{\perp}$.

Let the chemical potential $\zeta$ do not coincide with  $\varepsilon_d$. Such a situation usually occurs in Cd$_3$As$_2$.\cite{neupane,borisenko,jeon,liang} Consider the magnetization component $M_{\parallel}=\sum_iM_in_i$ directed along the magnetic field. At sufficiently high temperatures, when the de Haas -van Alphen oscillations are suppressed, the regime of weak magnetic fields, $H<H_T$, occurs, and $M_{\parallel}= \chi_{\parallel}H$ where $\chi_{\parallel}= \sum_{i,j}\chi^{ij}n_in_j$, and $\chi^{ij}$ is described by Eq.~(\ref{13}). At the same $H$ but at low temperatures, when $H_T$ decreases and $H_T\ll H$, the expression for $M_{\parallel}(H)$ follows from Eq.~(\ref{19}). As a result, we obtain for the difference of the {\it total} magnetizations at low and high temperatures,
 \begin{eqnarray}\label{26}
M_{\parallel,tot}(H)-\chi_{\parallel,tot} H=2C_0\cdot F(\theta)\cdot g(u),
 \end{eqnarray}
where the factor $2$ is due to existence of the two Dirac points in Cd$_3$As$_2$, $C_0$ is the constant factor depending on the parameters of the Dirac point,
\begin{equation}\label{27}
C_0=\frac{e(\zeta-\varepsilon_d)^2\sqrt{b_{zz}}}{12\pi^2\hbar^2 c (b_{zz}-a_z^2)},
\end{equation}
the factor $F(\theta)$ specifies the angular dependence of this difference,
\begin{eqnarray}\label{28}
F(\theta)&=&(\cos^2\theta+\gamma\sin^2\theta)^{1/2},\\
\gamma&\equiv& (b_{zz}-a_z^2)/A^2, \nonumber
\end{eqnarray}
$\gamma >0$ is the parameter characterizing the anisotropy of the Dirac cone, the function $g(u)$, Eq.~(\ref{20}), describes the oscillations of the magnetization with changing $H$, and $u$ is given by Eq.~(\ref{17}),
\begin{equation}\label{29}
u=\frac{c(\zeta-\varepsilon_d)^2b_{zz}}{2e\hbar A^2(b_{zz}-a_z^2)HF(\theta)}.
\end{equation}
Note that $u$ depends on the direction of the magnetic field through the same factor $F(\theta)$. This factor in $u$ specifies the angular dependence of extremal cross section of the Fermi surface. Interestingly, if one replaces the right hand side of Eq.~(\ref{26}) by $C\cdot [g(u_+)+g(u_-)]$ where $C$ and $u_{\pm}$ are given by formulas (\ref{22}), the obtained expression can be used even in the case when $|\zeta- \varepsilon_d|$ is so large that the electron spectrum begins to deviate from that given by Eq.~(\ref{25}).

Consider now the magnetic torque $K$ (per unit volume) of the sample at high and low temperatures. The value of this $K$ for Cd$_3$As$_2$  is independent of the direction of the two-dimensional vector $(n_x,n_y)$ in the $x$-$y$ plane, and so without the loss in generality, we may assume that the magnetic field lies in the $x$-$z$ plane, ${\bf n}=(\sin\theta,0,\cos\theta)$. Then, we obtain the following formula for the difference of the total magnetic torques at low ($K_{tot}^l$) and high ($K_{tot}^h$) temperatures:
 \begin{eqnarray}\label{30}
K_{tot}^l(H)\!-\!K_{tot}^h(H)\!=\!2C_0H\!\cdot\! \frac{(1-\gamma)\sin\theta\cos\theta}{F(\theta)}\!\cdot\! g(u)\,,
 \end{eqnarray}
where the constants $C_0$, $\gamma$, and the functions $F(\theta)$, $g(u)$ are the same as in Eq.~(\ref{26}).

Interestingly, a dependence of the magnetic susceptibility of liquid alloys Na$_{1-x}$Bi$_x$ on the concentration $x$ was measured many years ago,\cite{rob} and a noticeable diamagnetic deep on a smooth background was observed at the concentration $x=0.25$ which corresponds to the stoichiometric formula Na$_3$Bi. This result can be qualitatively understood from our Eq.~(\ref{13}) if one considers the $x$-dependence of the susceptibility near the point $x=0.25$ as the dependence of $\chi$ on the chemical potential in Na$_3$Bi.

\section{Line node semimetals}\label{line}

The band-contact lines are widespread in crystals, \cite{herring,m-sh14,kim,fang} e.g., they exist in graphite \cite{graphite}, and beryllium \cite{beryl}. However, as was mentioned in the Introduction, the band-contact lines in the line node semimetals are characterized by a relatively small difference $\varepsilon_{max} -\varepsilon_{min}\equiv 2\Delta$ of the maximum and minimum band-degeneracy energies as compared to the inherent energy scale of  crystals ($1-10$ eV). Such line nodes exist in  rhombohedral multilayer graphene  \cite{hei1,pie} and three-dimensional graphene networks. \cite{weng} Besides, the class of the topological line node semimetals includes Ca$_3$P$_2$, \cite{xie}  Cu$_3$NPd, \cite{kim,yu} CaAgP, \cite{yama} ZrSiS, \cite{schoop} PbTaSe$_2$, \cite{bian} and SrIrO$_3$. \cite{chen}

In most of these line node semimetals the spin-orbit interaction
is weak, and the band-contact lines exist if one neglects this interaction. The spin-orbit coupling lifts the degeneracy of the conduction and valence bands along the line, and a small gap between the bands appears. The effect of this gap on the magnetic susceptibility was studied in Refs.~\onlinecite{m-sv} and \onlinecite{m-sh}, and it was found that this effect, as a rule, is negligible. Because of this, we neglect the gap below. However, we shall point out a situation for which the effect of the gap becomes noticeable.

\subsection{Spectrum}\label{IIIa}

In the vicinity of a band-contact line along which the conduction and valance bands touch, let us introduce orthogonal curvilinear coordinates so that the axis ``$3$'' coincides with the line. The axes ``$1$'' and ``$2$'' are perpendicular to the third axis at every point of the band-contact line, and the appropriate coordinate $p_1$ and $p_2$ are measured from this line. In these coordinates, in the vicinity of the line, the most general form of the electron spectrum for the conduction and valence bands looks like
\begin{eqnarray}\label{31}
 \varepsilon_{c,v}\!\!&=&\!\varepsilon_d(p_3)\!+\!{\bf a}_{\perp}{\bf p}_{\perp}\pm E_{c,v},\\
 E_{c,v}^2\!\!&=&\!b_{11}p_1^2+b_{22}p_2^2, \nonumber
 \end{eqnarray}
where $\varepsilon_d(p_3)$ describes a dependence of the  degeneracy energy along the line; ${\bf p}_{\perp}=(p_1,p_2,0)$ and ${\bf a}_{\perp}=(a_1,a_2,0)$ are the vectors perpendicular to the line; the parameters of the spectrum  $b_{11}$, $b_{11}$, and ${\bf a}_{\perp}$ generally depend on $p_3$. As in Sec.~\ref{spectr}, it is implied here that the directions of the axes ``1'' and ``2'' are chosen so that the quadratic form $E_{c,v}^2$ is diagonal. Below we shall consider only the case when the length of the vector $\tilde{\bf a}_{\perp}\equiv (a_1/\sqrt{b_{11}},a_2/\sqrt{b_{22}},0)$ is less than unity,
\begin{eqnarray}\label{32}
 \tilde a_{\perp}^2=\frac{a_1^2}{b_{11}}+\frac{a_2^2}{b_{22}}<1,
 \end{eqnarray}
since at $\tilde a_{\perp}^2>1$ the giant anomaly of the magnetic susceptibility is absent. \cite{m-sv}

\begin{figure}[tbp] % %%%%%%%%%%%%%%%%%%%%%%%%%%%%%%%%%%%%%
 \centering  \vspace{+9 pt}
\includegraphics[scale=.95]{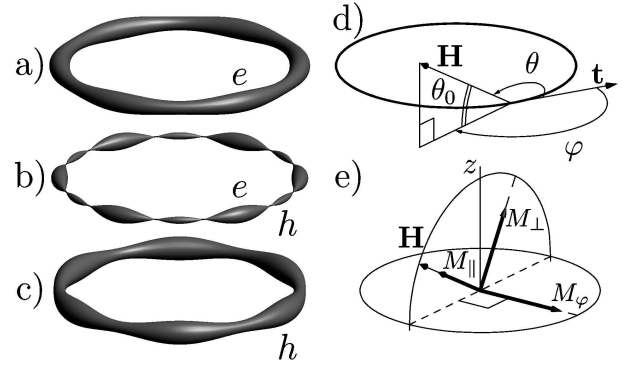}
\caption{\label{fig3} The Fermi surface of electrons in a line node semimetal at $\zeta>\varepsilon_d^0+\Delta$ (a), $\varepsilon_d^0+ \Delta>\zeta>\varepsilon_d^0-\Delta$ (b), and $\varepsilon_d^0- \Delta>\zeta$ (c). Letters e and h indicate the electron or hole type of the Fermi surface. (d) The band-contact line, and the definition of the angles $\theta$, $\theta_0$, and $\varphi$; the vector ${\bf t}$ is the tangent to the line at a point $p_3$. (e) Directions of the magnetization components $M_{\parallel}$, $M_{\perp}$, and $M_{\phi}$ with respect to ${\bf H}$ and the normal ${\bf z}$ to the plane of the band-contact line.
 } \end{figure}

Let us introduce the notations:
\[ \varepsilon_d^0\equiv\frac{\varepsilon_{max}+\varepsilon_{min}}{2}, \ \ \ \ \Delta \equiv \frac{\varepsilon_{max}-\varepsilon_{min}}{2},
\]
where $\varepsilon_{max}$ and $\varepsilon_{min}$ are the maximum and minimum values of $\varepsilon_d(p_3)$ in the line. When $\Delta$ is relatively small and $\tilde a_{\perp}^2<1$, the Fermi surface $\varepsilon_{c,v}({\bf p}_{\perp},p_3)=\zeta$ in the semimetals is a narrow electron or hole tube for $\zeta-\varepsilon_d^0\gtrsim \Delta$ or $\zeta- \varepsilon_d^0 \lesssim -\Delta$, respectively. The band-contact line lies inside these tubes. If  $|\zeta- \varepsilon_d^0| <  \Delta$, the Fermi surface consist of the electron and hole parts  and has a self-intersecting shape, Fig.~\ref{fig3}. When the chemical potential $\zeta$ passes through the critical energies $\varepsilon_d^0\pm \Delta$, the electron topological transitions of $3\frac{1}{2}$ kind occur. \cite{m-sh14} We shall assume below that the transverse dimensions of the Fermi-surface tube, which are of the order $|\zeta-\varepsilon_d|/V$ ($V\sim \sqrt{b_{ii}}$ as in Sec.~\ref{spectr}), are essentially less than the characteristic  radius of curvature for the band-contact line. In this case  practically all electron orbits in the ${\bf p}$-space, which are intersections  of the Fermi surface with planes perpendicular to the magnetic field, are small and lie near the band-contact line. In other words, a small region in the ${\bf p}$-space determines the local  electron energy spectrum in the magnetic field almost for any point of the line. This spectrum can be found,\cite{m-sh} and it has the form:
 \begin{eqnarray}\label{33}
\varepsilon_{c,v}^l(p_3)=\varepsilon_{d}(p_3) \pm \!\left(\frac{e\hbar\alpha H|\cos\theta|}{c}l\right)^{1/2}\!,
 \end{eqnarray}
where $\alpha=\alpha(p_3)=2(b_{11}b_{22})^{1/2}(1-\tilde a_{\perp}^2)^{3/2}$; $l$ is a non-negative integer ($l=0$, $1$, \dots), and $\theta=\theta(p_3)$ is the angle between the direction of the magnetic field and the tangent ${\bf t}$ to the band-contact line at the point with the coordinate $p_3$, Fig.~\ref{fig3}. Formula (\ref{33}) fails only for those points of the line for which $\theta$ is close to $\pi/2$.\cite{m-sh}  However, these points do not lead to the giant anomaly in the susceptibility. \cite{m-sv}

   %%%%%%%%%%%%%%%%%%%%%%%%%%%%%%%%%%%%%%%%%%

In a number of the topological semimetals, the band-contact line is a closed curve lying in a plane that is perpendicular to an axis of $n$-fold symmetry. Therefore, the line remains invariant under the rotations through the angles $2\pi i/n$  about this axis. (Here $i=1,2,\dots,n$). Then, $\varepsilon_d(p_3)$ is a periodic function with the period $L/n$ where $L$ is the length of the band-contact line in the ${\bf p}$-space. Below, for the purposes of illustration, we shall use the following simple model dependence for $\varepsilon_d(p_3)$:
 \begin{equation}\label{34}
 \varepsilon_d(p_3)=\varepsilon_d^0+\Delta \cos(2\pi p_3n/L).
 \end{equation}
When $\Delta$ is very small, one may expect that $b_{11}$, $b_{22}$, and $\tilde a_{\perp}^2$ are almost constant along the band-contact line, and this closed curve is approximately a circle. Note also that if a band-contact line in a topological semimetal is not a closed curve but it begins and ends on the opposite faces of the Brillouin zone, formula (\ref{34}) with $n=1$ provides a simple model for the function $\varepsilon_d(p_3)$ in this case as well.

\subsection{Magnetic susceptibility}\label{IIIb}

As in the case of the Weyl and Dirac semimetals, the total magnetic susceptibility of the line node semimetals consists of its special part determined by the electron states located near the band-contact line and a practically constant  background term specified by electron states located far away from this line. It is the special part that is responsible for dependences of the susceptibility on the magnetic field, temperature, and the chemical potential $\zeta$ when this $\zeta$ lies inside or close to the narrow energy interval from $\varepsilon_d^0- \Delta$ to $\varepsilon_d^0 +\Delta$.

In weak magnetic fields $H\ll H_T$, when $\Delta\varepsilon_H\ll T$, the special part of the magnetic susceptibility is practically  independent of $H$, whereas at $H>H_T$, a noticeable $H$-dependence of the susceptibility appears, and the magnetization becomes a nonlinear function of $H$. The boundary $H_T$ between the regions of weak and strong magnetic fields, which is defined by the condition  $\Delta\varepsilon_H \sim T$, is still estimated by Eq.~(\ref{10}) where $V$ is the characteristic slope of the Dirac cone, $V\sim (b_{ii})^{1/2}$. However, since in topological semimetals the value of $\Delta$ is relatively small, it is necessary to take into account an interrelations between the three parameters $T$, $\Delta\varepsilon_H$, and $\Delta$ when one analyzes the magnetic susceptibility and the magnetization. This situation differs from that of Refs.~\onlinecite{m-sv} and \onlinecite{m-sh} where $\Delta$ was assumed to be much larger than the temperature $T$ and the spacing between the Landau subbands $\Delta\varepsilon_H$.

In this section we present formulas for the susceptibility and the magnetization assuming the two-fold degeneracy of the conduction and valence bands in spin. However, in the case of a noncentrosymmetric line node semimetal with a strong spin-orbit interaction (e.g., in  PbTaSe$_2$ \cite{bian}), this degeneracy is absent. In this situation the formulas given below should be divided by two.

\subsubsection{Weak magnetic fields}

In weak magnetic fields when $\Delta\varepsilon_H\ll T$, we find the following expression for the longitudinal magnetic susceptibility $\chi_{\parallel}$ defining the magnetization component $M_{\parallel}=\chi_{\parallel}H$ parallel to the magnetic field:
\begin{eqnarray}\label{35}
 \chi_{\parallel}\!=\!\frac{e^2}{6\pi^2\!\hbar c^2}\!
 \!\!\int_{0}^{L}\!\!\!\!\!dp_3(b_{11}b_{22})^{\!1\!/2}\!(1-\tilde a_{\perp}^2)^{3\!/2}\!f'(\varepsilon_d)\cos^2\!\theta,
 \end{eqnarray}
where $L$ is the length of the band-contact line in the Brillouin zone; the integration is carried out over this line; $f'(\varepsilon_d)$ is the derivative of the Fermi function (\ref{12}),
\begin{equation}\label{36}
 f'(\varepsilon_d)=-\left[4T\cosh^2\left(\frac{\varepsilon_d(p_3)- \zeta}{2T}\right)\right]^{-1},
 \end{equation}
and $\theta=\theta(p_3)$ is the angle between the direction of the magnetic field and the tangent to the band-contact line at the point with the coordinate $p_3$. Formula (\ref{35}) describes the special part of the magnetic susceptibility.

At $T\ll 2\Delta$, we find from Eq.~(\ref{35}) that $\chi_{\parallel}(\zeta) =0$ if $\zeta$ does not lie between $\varepsilon_d^0-\Delta$ and  $\varepsilon_d^0+\Delta$. If  $|\zeta-\varepsilon_d^0|<\Delta$, we obtain
\begin{eqnarray}\label{37}
 \chi_{\parallel}(\zeta)\!=\!-\frac{e^2}{6\pi^2\!\hbar c^2}
\sum_{j}(b_{11}b_{22})^{\!1\!/2}\!(1-\tilde a_{\perp}^2)^{3\!/2}\! \frac{\cos^2\theta}{|d\varepsilon_d/dp_3|},
 \end{eqnarray}
where all the quantities in the right hand side of the formula are calculated at the points  $p_3=p_{3j}$. These $p_{3j}$ are found from the equation
\begin{equation}\label{38}
 \varepsilon_d(p_{3j})=\zeta.
 \end{equation}
Taking into account that $\varepsilon_d$, $b_{ii}$, and $\tilde a_{\perp}^2$ are periodic functions along the band-contact line, it can be shown that the sum in Eq.~(\ref{37}) reduces to the sum over $p_{3j}$ lying inside a single period of $\varepsilon_d(p_3)$, with  the additional factor $n/2$ appearing before the sum, and $\cos^2\theta$ being replaced by $\cos^2\theta_0$. Here $\theta_0$ is the angle between the magnetic field and the plane of the line, Fig.~\ref{fig3}.

\begin{figure}[tbp] % %%%%%%%%%%%%%%%%%%%%%%%%%%%%%%%%%%%%%
 \centering  \vspace{+9 pt}
\includegraphics[scale=.99]{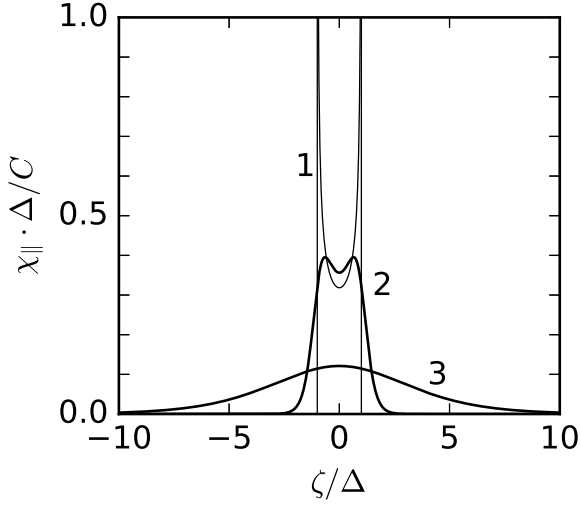}
\caption{\label{fig4} The dependence of $\chi_{\parallel}$ on the chemical potential $\zeta$ at 1) $T\to 0$, Eq.~(\ref{39}) 2) $T/\Delta=0.25$ and 3) $T/\Delta=2$, Eqs.~(\ref{39}), (\ref{41}). At $T/\Delta\ge 2$, the $\zeta$-dependences of the susceptibility calculated with Eqs.~(\ref{39}), (\ref{41}) and with approximate formula (\ref{42}) practically coincide. The susceptibility is measured in units of $C/\Delta$ where the constant $C$ is defined by formula (\ref{43}); $\zeta$ is measured from $\varepsilon_d^0$.
 } \end{figure}

If $\varepsilon_d(p_3)$ is given by Eq.~(\ref{34}), one finds the two   points $p_{3j}$ ($j=1,2$) inside the period:
\[
p_{3j}=\pm \frac{L}{2\pi n} \arccos\left(\frac{\zeta-\varepsilon_d^0}{\Delta}\right),
\]
and the following derivative $|d\varepsilon_d/dp_3|$ which is identical for both the points $p_{3j}$:
\[
 \left|\frac{d\varepsilon_d}{dp_3}\right|=\frac{2\pi n}{L}\sqrt{\Delta^2-(\zeta-\varepsilon_d^0)^2}.
\]
Eventually, in the case of Eq.~(\ref{34}) we obtain,
\begin{eqnarray}\label{39}
 \chi_{\parallel}(\zeta)\!=\!-\frac{e^2}{6\pi^2\!\hbar c^2}
\frac{L(b_{11}b_{22})^{1\!/2}\!(1-\tilde a_{\perp}^2)^{3\!/2}\!\cos^2\theta_0} {2\pi \sqrt{\Delta^2-(\zeta-\varepsilon_d^0)^2}},
 \end{eqnarray}
where $b_{ii}$ and $\tilde a_{\perp}^2$ are calculated at one of $p_{3j}$. This function $\chi_{\parallel}(\zeta)$ is shown in Fig.~\ref{fig4} assuming that $b_{ii}$ and $\tilde a_{\perp}^2$ are constant along the line. When $\zeta$ tends, e.g., to  $\varepsilon_{max}= \varepsilon_d^0 +\Delta$, Eq.~(\ref{38}) reduces to $\zeta= \varepsilon_{max}-Bp_{3j}^2$ where
 \begin{eqnarray}\label{40}
 B=\frac{2\pi^2 n^2 \Delta}{L^2},
 \end{eqnarray}
and one derives $\chi_{\parallel}\propto (\varepsilon_{max}- \zeta)^{-1/2}$ from Eq.~(\ref{39}), with the coefficient of proportionality agreeing with Eq.~(12) from Ref.~\onlinecite{m-sv}. Of course, at $|\varepsilon_{max} -\zeta|\lesssim T$ the divergence of $\chi_{\parallel}$ in Eq.~(\ref{39}) is cut off as in the case of the Dirac points. Thus, formulas (\ref{37}) and (\ref{39}) extend the result of Ref.~\onlinecite{m-sv} for the magnetic susceptibility near the point of the electron topological transition to the whole interval from $\varepsilon_{min}$ to $\varepsilon_{max}$. In the middle of this interval the divergence of $\chi_{\parallel}$ is absent, but the susceptibility is still large due to a small value of $\Delta$ in the denominator of Eq.~(\ref{39}) and essentially depends on $\zeta$.

   %%%%%%%%%%%%%%%%%%%%%%%%%%%%%%%%%%%%%%%%%%

In a number of line node semimetals the degeneracy of the conduction and valence bands along the line is lifted by the spin-orbit interaction which generates a small gap $\Delta_{so}(p_3)$ between the bands. If the temperature is lower than this gap, the divergence of the susceptibility at $\zeta\to \varepsilon_{max}$ in Eq.~(\ref{39}) is cut off at $|\varepsilon_{max} -\zeta|\lesssim \Delta_{so}(p_{3i})$. In other words, at low $T$ the magnitude of the susceptibility near the critical energies $\varepsilon_{max}$ and $\varepsilon_{min}$ is determined by the spin-orbit gap rather than by the temperature. The  detailed analysis of the magnetic susceptibility at the chemical potential lying in this gap is presented in Ref.~\onlinecite{m-sh}.

Consider now the case when the temperature is not small as compared with $\Delta$ (and $T\gg \Delta\varepsilon_H$). In this situation, it is convenient to rewrite formula (\ref{35}) as follows:
 \begin{equation} \label{41}
 \chi_{\parallel}(\zeta,T)=- \int_{\varepsilon_d^0-\Delta}^{\varepsilon_d^0+\Delta}d\varepsilon \chi_{\parallel}(\varepsilon,0)f'(\varepsilon),
 \end{equation}
where $\chi_{\parallel}(\zeta,0)$ is given by Eq.~(\ref{37}) or Eq.~(\ref{39}). The $\zeta$-dependence of this $\chi_{\parallel}(\zeta,T)$ is shown in Fig.~\ref{fig4}, assuming that $b_{ii}$ and $\tilde a_{\perp}^2$ are independent of $p_3$. In the limiting case when $\Delta$ is so small (or the temperature is so high) that $\Delta\ll T$ (the interrelation between $\Delta$ and $\Delta\varepsilon_H$ may be arbitrary), an explicit formula for $\chi_{\parallel}(\zeta,T)$ can be obtained. In this case $\varepsilon_d(p_3)$ is approximately equal to $\varepsilon_d^0$ in formula (\ref{36}), and the factor $(b_{11}b_{22})^{\!1\!/2}\!(1-\tilde a_{\perp}^2)^{3\!/2}$ in Eq.~(\ref{35}) can be replaced by its value averaged over the band-contact line. Taking into account that the band-contact line is invariant under the rotations through the angles $2\pi i/n$, it can be shown that $\int_0^L\cos^2\theta dp_3=(L\cos^2\theta_0)/2$. Eventually, we arrive at
\begin{eqnarray}\label{42}
 \chi_{\parallel}(\zeta,T)\!=\!-\frac{e^2\cos^2\theta_0}{6\pi^2\!\hbar c^2}\cdot \frac{\int_{0}^{L}\!\!dp_3(b_{11}b_{22})^{\!1\!/2}\!(1-\tilde a_{\perp}^2)^{3\!/2}}{8T\cosh^2[(\varepsilon_d^0- \zeta)/2T]}.
 \end{eqnarray}
The data of Fig.~\ref{fig4} show that this formula well describes $\chi_{\parallel}(\zeta,T)$ at $T/\Delta \ge 2$.

Interestingly, formula (\ref{42}) can be also obtained from Eq.~(\ref{41}) if the susceptibility at zero temperature  $\chi_{\parallel}(\varepsilon,0)$  is replaced by $C\delta(\varepsilon-\varepsilon_d^0)$, where the constant $C$ is
\begin{eqnarray}\label{43}
C&=&\int_{-\Delta}^{\Delta}\chi_{\parallel}(\zeta)d(\zeta-\varepsilon_d^0) \nonumber \\ &=&-\frac{e^2 L}{12\pi^2\!\hbar c^2}
(b_{11}b_{22})^{1\!/2}\!(1-\tilde a_{\perp}^2)^{3\!/2}\!\cos^2\theta_0,
\end{eqnarray}
and $\chi_{\parallel}(\zeta)$ is given by Eq.~(\ref{39}). Formula
$\chi_{\parallel}(\zeta,0)=C\delta(\zeta-\varepsilon_d^0)$ corresponds to the result of Koshino. \cite{kosh15} Therefore, at $T\ge 2\Delta$, i.e., when the structure of $\chi_{\parallel}(\zeta,0)$ shown in Fig.~\ref{fig4} at $|\zeta-\varepsilon_d^0|<\Delta$ becomes unimportant, Koshino's result is equivalent to ours.

It is necessary to emphasize that the susceptibility  $\chi_{\parallel}$ found in this section is invariant under rotations of the magnetic field through any angle $\phi$ about the $z$ axis that is perpendicular to  the plane of the band-contact line. This $\chi_{\parallel}$ is proportional to $\cos^2\theta_0$ where $\theta_0$ is the angle between the magnetic field and this plane. When $\theta_0$ differs from zero and $\pi/2$, there is also a nonzero component $M_{\perp}$ of the magnetization that is perpendicular to the magnetic field ${\bf H}$, Fig.~\ref{fig3}. This component lies in the plane passing through the vector ${\bf H}$ and the $z$ axis. Interestingly, the magnetic susceptibility $\chi_{\perp}$ defined by the relation $M_{\perp}=\chi_{\perp}H$ is also described by formulas (\ref{39}) and (\ref{42}) in which $\cos^2\theta_0$ should be replaced by $\cos\theta_0\sin\theta_0$. This $\chi_{\perp}$ determines the magnetic torque  $K_{\phi}=\chi_{\perp}H^2$. As to the background contributions to the total susceptibilities $\chi_{\parallel,tot}$ and $\chi_{\perp,tot}$, they look like $\chi_{\parallel}^0 = \chi_{zz}^0 +(\chi_{x-y}^0-\chi_{zz}^0) \cos^2\theta_0$, $\chi_{\perp}^0= \sin\theta_0\cos\theta_0(\chi_{x-y}^0 - \chi_{zz}^0)$  where $\chi_{x-y}^0$ and $\chi_{zz}^0$ are the components of the background susceptibility tensor in the plane of the line and perpendicular to this plane, respectively. Thus, in weak magnetic fields the angular dependences of the special and background contributions to the magnetic susceptibility have the same form.

\subsubsection{Strong magnetic fields}\label{IIIb2}

Generalizing the results of Ref.~\onlinecite{m-sh}, we find the following expression for the magnetization in strong magnetic fields ($\Delta\varepsilon_H\gg T$):
\begin{eqnarray}\label{44}
 M_{\parallel}(\zeta,H)\!=\!\frac{e^{3/2}H^{1/2}} {2\pi^2\hbar^{3/2}c^{3/2}}\!\!\int_{0}^{L}\!\!\!\!\!dp_3 |\cos\theta|^{3/2}\!\sqrt{\alpha(p_3)} K(u),
 \end{eqnarray}
 where
\begin{eqnarray*}
 \alpha(p_3)\!\!&=&\!\!2(b_{11}b_{22})^{\!1\!/2}\!(1-\tilde a_{\perp}^2)^{3\!/2}, \\
 K(u)\!\!\!&=&\!\!\! \frac{3}{2}\zeta(-\frac{1}{2},\![u]\!+\!1\!)+\sqrt{u}([u]+\frac{1}{2}). \nonumber
 \end{eqnarray*}
$\zeta(x,a)$ is the Hurwitz zeta function, $[u]$ is the integer part of $u$,
\begin{equation}\label{45}
u=\frac{[\zeta-\varepsilon_d(p_3)]^2 c}{e\hbar  \alpha(p_3) H|\cos\theta|}=\frac{cS(p_3)}{2\pi e\hbar H},
\end{equation}
$S(p_3)$ is the area of the cross section of the Fermi surface by the plane perpendicular to the magnetic field and passing through the point $p_3$. The quantity $u$ is similar to that defined by Eq.~(\ref{17}). In calculations with Eqs.~(\ref{44}) and (\ref{45}), it is convenient to use the geometrical relation  $\cos\theta= \cos\theta_0\cos\varphi$ in which $\varphi$ is the angle between the tangent to the band-contact line at the point with the coordinate $p_3$ and the projection of magnetic field on the plane of the line, and $\theta_0$ is the angle between the magnetic field and this plane, Fig.~\ref{fig3}. It is clear that only the angle $\varphi$ depends on $p_3$.

As in the case of the weak magnetic fields, the magnetization in the region of strong magnetic fields has the component ${\bf M}_{\perp}$ that is perpendicular to ${\bf H}$ and lies in the plane containing the vectors ${\bf H}$ and ${\bf z}$. This component is expressed via $M_{\parallel}$ as follows:
\begin{eqnarray}\label{46}
M_{\perp}= \tan\theta_0 M_{\parallel}.
 \end{eqnarray}
However, in contrast to the case of weak fields, in the region of strong magnetic fields there is also a nonzero component ${\bf M}_{\phi}$ directed perpendicularly both to ${\bf H}$ and ${\bf M}_{\perp}$, Fig.~\ref{fig3}. This component is described by the formula that is similar to Eq.~(\ref{44}):
\begin{eqnarray} \label{47}
 M_{\phi}(\zeta,H)\!=\!\frac{e^{3/2}H^{1/2}(\cos\theta_0)^{1/2}} {2\pi^2\hbar^{3/2}c^{3/2}}\!\!\int_{0}^{L}\!\!\!\!\!dp_3 \sigma(\varphi)\sin\varphi \nonumber \\
 \times \sqrt{\alpha(p_3)|\cos\varphi|} K(u),
 \end{eqnarray}
where $\sigma(\varphi)$ is a sign of $\cos\varphi$. The components $M_{\perp}$ and $M_{\phi}$ determines the magnetic torque of the sample.

Strictly speaking, formulas (\ref{44}) and (\ref{47}) describe $M_{\parallel}$ and $M_{\phi}$ at $T \to 0$. For nonzero temperatures  (including the case of weak magnetic fields, $T\gg \Delta \varepsilon_H$), the magnetizations $M_{\parallel}(\zeta,H,T)$, $M_{\perp}(\zeta,H,T)$ and $M_{\phi}(\zeta,H,T)$ can be calculated with the relationship: \cite{rum}
 \begin{equation}\label{48}
 M_{\parallel,\perp,\phi}(\zeta,H,T)=-\int_{\!-\infty}^{\infty}\!\! d\varepsilon M_{\parallel,\perp,\phi}(\varepsilon,H,0)f'(\varepsilon),
 \end{equation}
where $f'(\varepsilon)$, $M_{\parallel,\perp,\phi}(\zeta,H,0)$ are given by Eqs.~(\ref{36}), (\ref{44}), (\ref{46}) and (\ref{47}), respectively.

Consider now $M_{\parallel}(\zeta,H)$ and $M_{\phi}(\zeta,H)$ in limiting cases. When $\Delta\varepsilon_H\ll \Delta$, the results of Ref.~\onlinecite{m-sh} are valid. In particular,
when $\zeta$ is not close to the critical energies $\varepsilon_d^0\pm \Delta$, the magnetization $M_{\parallel}$ is described by usual formulas \cite{Sh} for the de Haas van Alphen effect, with the phase of the oscillations being shifted by $\pi$. \cite{shen} As in Sec.~\ref{IIb}, this shift is caused by the large value of the $g$ factor, $g=2m/m_*$, occurring even at a weak spin-orbit interaction. \cite{jetp} Here $m_*$ is the cyclotron mass and $m$ is the electron mass. This large value of $g$ factor is due to the Berry phase $\pi$ for the electron orbits surrounding the band-contact lines. \cite{prl,shen}

When $\zeta$ tends to one of the critical energies $\varepsilon_d^0\pm \Delta$, the magnetization $M_{\parallel}$ is determined by those critical points in the band-contact line which correspond to this energy. For such a point, we obtain  $\varepsilon_d(p_3)-\zeta\approx Bp_3^2$ where $B$ is a constant [see, e.g., Eq.~(\ref{40})], and  now $u\propto p_3^4/(H|\cos\theta|)$. Changing the variable of the integration in Eq.~(\ref{44}) from $p_3$ to $u$, we arrive at
\begin{eqnarray}\label{49}
 M_{\parallel}(H)\!=\!-\frac{f_0e^{7/4}\alpha_c^{3/4}n} {\pi^2\hbar^{5/4}c^{7/4}|B|^{1/2}}\cdot H^{3/4}(\cos\theta_0)^{7/4}\Phi_n(\phi),
 \end{eqnarray}
where $f_0\approx 0.156$, $\alpha_c$ denotes the value of $\alpha$ at one of the critical points, $\phi$ is the angle between the tangent to the band-contact line at this point and the projection of magnetic field on the plane of the line, the factor $\Phi_n(\phi)$,
\begin{eqnarray}\label{50}
\Phi_n(\phi)=\frac{1}{n}\sum_{i=1}^n \left|\cos\left(\phi+\frac{2\pi i}{n}\right)\right|^{7/4}\!\!\!,
 \end{eqnarray}
describes the dependence of $M_{\parallel}(H)$ on the direction of this projection. Formula (\ref{49}) shows that $M_{\parallel}\propto H^{3/4}$ and  that the angular dependence of the magnetization on $\theta_0$ has changed as compared to the case of the weak magnetic fields. Moreover, $M_{\parallel}(H)$ is not isotropic in the plane of the band-contact line since the functions $\Phi_n(\phi)$ are not constants. However, when ${\bf H}$ rotates about the $z$ axis, the variation of $M_{\parallel}$ is relatively small. This follows from that the functions $\Phi_n(\phi)$ are well approximated by the expressions:
\begin{eqnarray}\label{51}
\Phi_4(\phi)&\approx& 0.5526-0.0226\cos(4\phi), \\ \Phi_3(\phi)&=&\Phi_6(\phi)\approx 0.5249+0.0066\cos(6\phi). \nonumber
\end{eqnarray}
and their change with $\phi$ is of the order of 1-4 per cent.

Interestingly, at $\Delta\varepsilon_H\ll \Delta$ and $\zeta\to \varepsilon_d^0\pm \Delta$ there is also a nonzero component ${\bf M}_{\phi}$. This component is given by Eq.~(\ref{49}), with $(\cos\theta_0)^{7/4}$ and $\Phi_n(\phi)$ being replaced by $(\cos\theta_0)^{3/4}$ and $\Psi_n(\phi)$, respectively. Here
\begin{equation}\label{52}
 \Psi_n(\phi)\!=\!\frac{1}{n}\sum_{i=1}^n\sigma_i \sin\!\left(\!\phi+\frac{2\pi i}{n}\right)\! \left|\cos\!\left(\!\phi+\frac{2\pi i}{n}\right)\right|^{3/4}\!\!\!\!\!,
\end{equation}
and $\sigma_i$ is a sign of $\cos\left(\phi+\frac{2\pi i}{n}\right)$. The functions $\Psi_n(\phi)$ can be approximately described by the expressions:
\begin{eqnarray}\label{53}
\Psi_4(\phi)&\approx&\!-0.046\sin(4\phi)-0.01\sin(8\phi), \\ \Psi_3(\phi)&=&\Phi_6(\phi)\approx 0.02\sin(6\phi)-0.0045\sin(12\phi). \nonumber
\end{eqnarray}
Thus, in the region of the magnetic fields, $T\ll \Delta\varepsilon_H\ll \Delta$, the component $M_{\phi}$ at $\zeta\to \varepsilon_d^0\pm \Delta$ is an order of magnitude  smaller than $M_{\parallel}$.

\begin{figure}[tbp] % %%%%%%%%%%%%%%%%%%%%%%%%%%%%%%%%%%%%%
 \centering  \vspace{+9 pt}
\includegraphics[scale=.99]{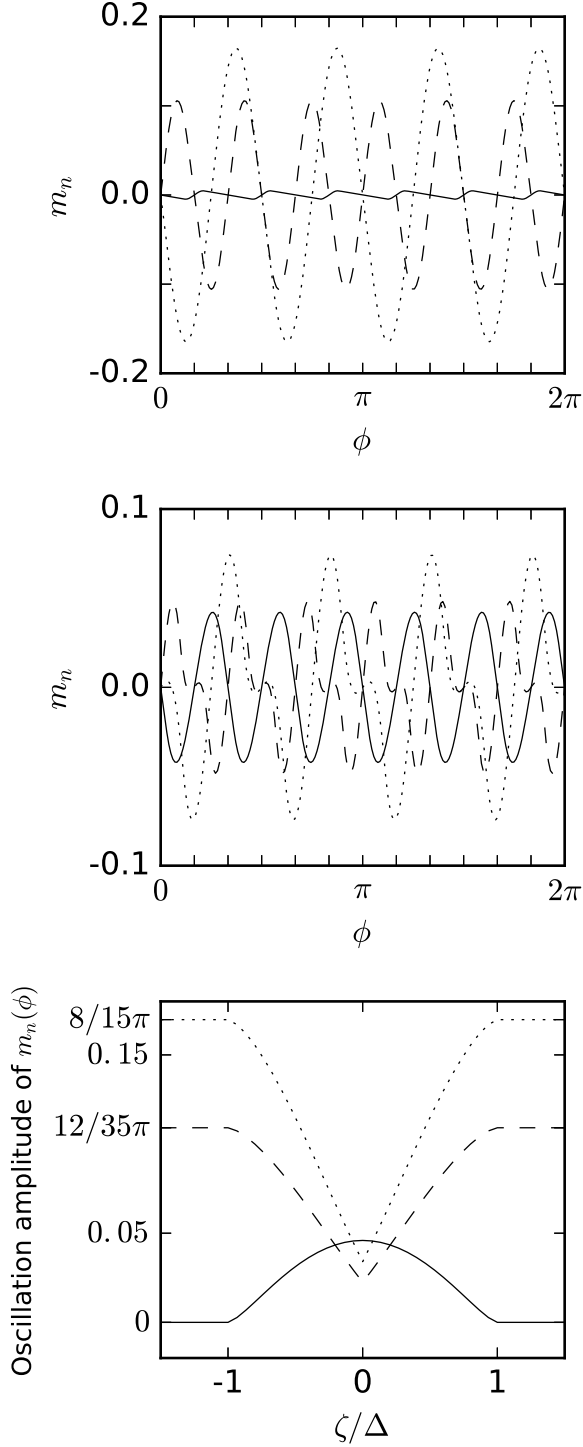}
\caption{\label{fig5} Dependences $m_n(\phi)$, Eq.~(\ref{60}), at $\tilde\zeta=0.9$ (top) and $0.25$ (middle). The bottom panel shows the $\zeta$-dependence of the amplitude of the oscillations in $m_n(\phi)$. Here $\zeta$ is measured from $\varepsilon_d^0$, the solid lines correspond to $n=3$, the dashed lines to $n=6$, and the dotted lines to $n=4$.
 } \end{figure}   %%%%%%%%%%%%%%%%%%%%%%%%%%%%%%%%%%%%%%%%%%

Consider now the case of strong magnetic fields ($T\ll \Delta\varepsilon_H$), and let $\Delta$ be so small (or the magnetic field be so large) that $\Delta\ll \Delta\varepsilon_H$. If $|\zeta- \varepsilon_d^0|\gg \Delta\varepsilon_H$, the de Haas - van Alphen oscillations occur,  with the phase of the oscillations being shifted by $\pi$. If $|\zeta- \varepsilon_d^0|\lesssim \Delta$, the parameter $u$ in Eq.~(\ref{45}) is small ($u\ll 1$) practically for all points of the band-contact line, and Eq.~(\ref{44}) reduces to
\begin{eqnarray}\label{54}
 M_{\parallel}(\zeta,H)\!\!&\approx&\!\frac{3e^{3/2}H^{1/2} } {4\pi^2\hbar^{3/2}c^{3/2}} \zeta(-\frac{1}{2},1)(\cos\theta_0)^{3/2}G_{\parallel}(\phi) \nonumber \\
 &+&\!\frac{e\cos\theta_0}{4\pi^2\hbar^2 c}\!\!\int_{0}^{L}\!\!\!dp_3 |(\zeta-\varepsilon_d(p_3))\cos\varphi(p_3)|,
 \end{eqnarray}
where $\zeta(-\frac{1}{2},1)=-\zeta(3/2)/4\pi\approx -0.653/\pi$,  $\zeta(a)$ is the Riemann zeta function,
\begin{eqnarray}\label{55}
G_{\parallel}(\phi)=\int_{0}^{L}\!\!\!\!dp_3 |\!\cos\varphi(p_3)|^{3/2}\!\sqrt{\alpha(p_3)},
 \end{eqnarray}
$\varphi(p_3)$ is the angle between the tangent to the band-contact line at the point $p_3$ and the projection of magnetic field on the plane of the line, and the angle $\phi$ defines a direction of this projection in the plane. The second term in Eq.~(\ref{54}) is relatively small, and it gives only a correction to the first one. The first term is independent of $\zeta$ and reveals that in this field region, $M_{\parallel}\propto H^{1/2}$ rather than $M_{\parallel}\propto H^{3/4}$.

As to the component  $M_{\phi}$, we obtain from Eq.~(\ref{47}),
\begin{eqnarray} \label{56}
 M_{\phi}(\zeta,H)\!\!&\approx&\!\!\frac{3e^{3/2}H^{1/2} } {4\pi^2\hbar^{3/2}c^{3/2}} \zeta(-\frac{1}{2},1)(\cos\theta_0)^{1/2}G_{\phi}(\phi) \nonumber \\
 &+&\!\!\frac{e}{4\pi^2\hbar^2 c}\!\!\int_{0}^{L}\!\!\!dp_3 |(\zeta-\varepsilon_d(p_3))|\sigma(\varphi)\sin\varphi, \\
G_{\phi}(\phi)\!\!&=&\!\!\int_{0}^{L}\!\!\!\!dp_3 \sigma(\varphi)\sin\varphi  |\!\cos\varphi(p_3)|^{1/2}\!\sqrt{\alpha(p_3)}, \label{57}
  \end{eqnarray}
where $\sigma(\varphi)$ is a sign of $\cos\varphi$. Of course, Eqs.~(\ref{54}) and (\ref{56}) are true only when $\cos\theta_0$ is not small so that one has $u\ll 1$ at $\theta=\theta_0$ for the parameter $u$ defined by Eq.~(\ref{45}).

Let us apply the results of this section to the model in which
$\varepsilon_d(p_3)$ is described by Eq.~(\ref{34}) with a small value of $\Delta$, i.e., when $\Delta \ll \varepsilon_d^0$. In this case, in the leading order in the small parameter $\Delta/\varepsilon_d^0$, we may expect that $\alpha(p_3)\approx$ const., and the band-contact line is practically a circle. In this approximation, we find that the first (main) term $M_{\parallel}(H)$ in Eq.~(\ref{54}) is equal to
\begin{eqnarray}\label{58}
 M_{\parallel}(H)\!\!&\approx&\!-\frac{0.85}{\pi^4}\left(\frac{e} {\hbar c}\right)^{3/2}\!\!(\alpha H)^{1/2}(\cos\theta_0)^{3/2}L,
 \end{eqnarray}
and it is isotropic in the plane of the band-contact line. In the same approximation, the first term in Eq.~(\ref{56}) vanishes, and the component $M_{\phi}$ is determined by the second term which reduces to
\begin{eqnarray}\label{59}
M_{\phi}(\zeta,\!\phi)\!\!&=&\!\!\frac{eL\Delta}{4\pi^2\hbar^2 c}m_n(\tilde\zeta,\phi), \\
m_n(\tilde\zeta,\phi)\!\!&=&\!\!\frac{1}{2\pi}\!\!
\int_{0}^{2\pi}\!\!\!\!\!\!\!d\varphi |\tilde\zeta\!\!-\!\cos(n[\varphi+\phi])|\sigma(\varphi)\! \sin(\varphi),~~~ \label{60}
\end{eqnarray}
This $M_{\phi}$ is independent of the magnetic field magnitude, but it depends on $\tilde\zeta\equiv (\zeta-\varepsilon_d^0)/\Delta$ and the angle $\phi$ defining a direction of the magnetic field projection on the plane of the band-contact line. The angle $\phi$ in Eq.~(\ref{60}) is  measured from the tangent to this line at the point $p_3=0$. Formulas (\ref{59}), (\ref{60}) are valid when $\Delta, \zeta- \varepsilon_d^0 \ll \Delta\varepsilon_H \ll \varepsilon_d^0$.

At $\tilde\zeta\ge 1$ the quantities $m_n$ are independent of $\tilde\zeta$, and Eq.~(\ref{60}) yields the following simple expressions:
\begin{eqnarray}\label{61}
m_3=0; \ \
m_4=-\frac{8}{15\pi}\sin(4\phi); \ \
m_6=\frac{12}{35\pi}\sin(6\phi).
\end{eqnarray}
At $\tilde\zeta\le -1$, $m_n$ are still given by formulas (\ref{61}) but with the opposite sign. The functions $m_n(\phi)$ at two values of $\tilde\zeta < 1$ are shown in Fig.~\ref{fig5}. We see that with decreasing $\tilde\zeta$, the shape of the curves $m_4(\phi)$ and $m_6(\phi)$ deform, and at $\tilde\zeta=0$ we find that $m_4\propto \sin(8\phi)$ and $m_6\propto \sin(12\phi)$. On the other hand, the period of the function $m_3(\phi)$ remains constant on decreasing $|\tilde\zeta|$, but the amplitude and the shape of its oscillations change essentially, and $m_3(\phi)\propto \sin(6\phi)$ at $\tilde\zeta=0$. In Fig.~\ref{fig5} we also show how the amplitudes of the oscillations in $m_n(\phi)$ depend on $\tilde\zeta$.

Therefore, when $\Delta$ is so small that it is less than an  experimental value of $T$, and the chemical potential is close to $\varepsilon_d^0$, $|\zeta-\varepsilon_d^0|\lesssim \Delta$, the behavior of magnetization with increasing $H$ can be described as follows: In the weak magnetic fields ($\Delta\varepsilon_H<T$), one has $M_{\parallel}=\chi_{\parallel}H$, with $\chi_{\parallel}$ being  described by Eq.~(\ref{42}). In strong magnetic fields  ($\Delta\varepsilon_H>T$), we find $M_{\parallel}\propto H^{1/2}$; see Eq.~(\ref{58}). The intermediate asymptotic behavior $M_{\parallel}\propto H^{3/4}$ does not occur in this situation. Interestingly, the same dependences of  $\chi_{\parallel}$ on $\zeta$ in the weak magnetic fields \cite{mc,safran,kosh07} and of the magnetization on $H$ in the strong fields \cite{sharapov} were obtained for the case of a Dirac point in a two-dimensional layer in the magnetic field that is perpendicular to this layer. In other words, in the case of an ``ideal'' topological line node semimetal (when $\Delta\to 0$), its magnetization $M_{\parallel}$ is somewhat similar to the magnetization of graphene if one does not pay attention to the dependences of $M_{\parallel}$ on $\theta_0$ and $\phi$ in the semimetals. As to $M_{\phi}$, this quantity is zero in the weak magnetic fields, but in the region of the strong fields, its magnitude has a tendency to increase with $H$. Within the framework of Eq.~(\ref{34}), this magnitude is eventually saturated, and formulas (\ref{59}), (\ref{60}) give these saturated values of $M_{\phi}$.

If a band-contact line in a topological semimetal is not a closed curve but it begins and ends on the opposite faces of the Brillouin zone, the formulas of this section remain valid. However, the angular dependences of the magnetic susceptibility in the weak magnetic fields and of the magnetization in the strong magnetic fields change and are determined by the shape of the line.

\subsection{Example: Ca$_3$P$_2$}

\begin{figure}[tbp] % %%%%%%%%%%%%%%%%%%%%%%%%%%%%%%%%%%%%%
 \centering  \vspace{+9 pt}
\includegraphics[scale=1.0]{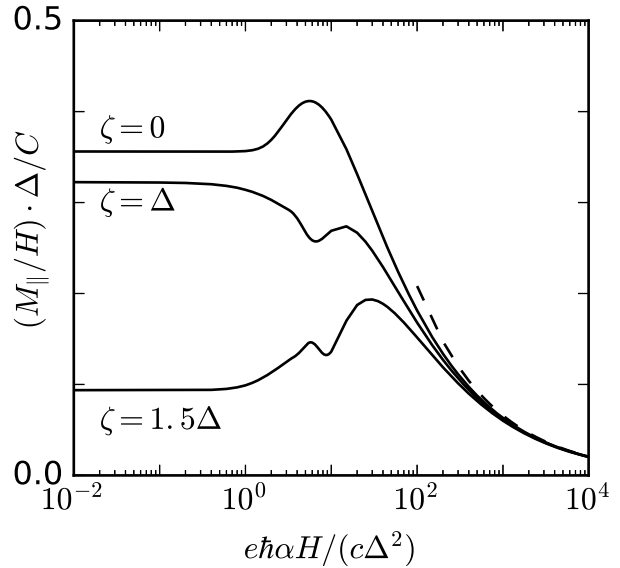}
\caption{\label{fig6} The $H$-dependence of the quantity  $M_{\parallel}/H$ calculated with formulas (\ref{44}), (\ref{48}), and (\ref{34}) at $\alpha=\,$const, $T=0.25\Delta$, $\theta_0=0$, $\phi=0$ for the three values of the chemical potential: $\zeta=0$, $\Delta$, and $1.5\Delta$ ($\zeta$ is measured from $\varepsilon_d^0$). The dashed line shows $M_{\parallel}/H$ according to Eq.~(\ref{58}). The quantity $M_{\parallel}/H$ is measured in units of $C/\Delta$ where the constant $C$ is defined by formula (\ref{43}); the $H$-axis is presented in the logarithmic scale.
 } \end{figure}   %%%%%%%%%%%%%%%%%%%%%%%%%%%%%%%%%%%%%%%%%%

In Ca$_3$P$_2$ the band-contact line looks like a circle which lies in the $x$-$y$ mirror-reflection plane and is perpendicular to a six-fold axis (the $z$ axis). \cite{xie} In this case the curvilinear coordinates introduced in Sec.~\ref{IIIa} coincide with cylindrical ones ($p_r,p_{\varphi},p_z$), and we have the following correspondence of the axes: $p_1=p_z$, $p_2=p_r$, $p_3=p_{\varphi}$. The spectrum near the band-contact line is described by Eq.~(\ref{31}) with ${\bf a}_{\perp}=(0,a_r,0)$ where $a_r$ is a constant. In principle, the parameters $b_{rr}$, $b_{zz}$, and $a_r$ can be found from the calculations of the electron-band structure of Ca$_3$P$_2$ presented in Ref.~\onlinecite{xie}, and as follows from Fig.~3 of that paper, the value $a_r$ is relatively small as compared to $\sqrt{b_{rr}}$, i.e., $\tilde a_{\perp}^2<1$. This means that the giant anomaly of the magnetic susceptibility should occur in Ca$_3$P$_2$, and formulas of Sec.~\ref{IIIb} enable one to calculate the magnetization and the magnetic susceptibility of this semimetal. Although the value of $\Delta$ is not extractable from the data presented by Xie et al.,\cite{xie} it is seen from their figures that $\Delta$ is relatively small, and so the reasonable assumption is that the model based on Eq.~(\ref{34}) with $n=6$ is applicable to Ca$_3$P$_2$. Therefore, the data of Figs.~\ref{fig4} and \ref{fig5} are likely to be suitable for this semimetal. Using this model and Eqs.~(\ref{44}) and (\ref{48}), we also calculate the $H$-dependence of the quantity $M_{\parallel}/\bar\chi_{\parallel}H$ for various positions of $\zeta$ relative to $\varepsilon_d^0$ at temperature $T=0.25\Delta$ and the magnetic fields lying in the plane of the band-contact line (i.e., at $\theta_0=0$), Fig.~\ref{fig6}. In figure~\ref{fig6}, $\bar\chi_{\parallel}=C/\Delta$, and $C$ is defined by Eq.~(\ref{43}). The quantity $M_{\parallel}/\bar\chi_{\parallel}H$ at $T=0.25\Delta$ is practically independent of the angle $\phi$, and for definiteness, we take $\phi=0$ in Fig.~\ref{fig6} (the angle $\phi$ is measured from the tangent to the band-contact line at the point $p_3=0$). The data of Fig.~\ref{fig6} show that at weak magnetic fields the values of  $M_{\parallel}/\bar\chi_{\parallel}H$ coincide with the results obtained from  Eq.~(\ref{35}), cf.\ Fig.~\ref{fig4}. At very strong magnetic fields, $\Delta\varepsilon_H\gg \Delta$, the magnetization is described by  formula (\ref{58}) and is independent of $\zeta$. At $\Delta\varepsilon_H\sim \Delta$, a de Haas -van Alphen oscillation is visible, and for $\zeta=\varepsilon_d^0+\Delta$, this oscillation is superimposed on the dependence approximately described by formula (\ref{49}). Thus, this figure demonstrates that $M_{\parallel}(H)/H$ is sensitive  to the position of the chemical potential relative to characteristic energies of the band-contact line.

\section{Conclusions}\label{conc}

We have considered the magnetic susceptibility and magnetization of electrons in topological Weyl, Dirac, and line node semimetals. In weak magnetic fields, when the spacing between Landau subbands $\Delta\varepsilon_H\propto \sqrt{H}$ is less than the temperature $T$, the susceptibility is independent of $H$. In the opposite case, when $\Delta\varepsilon_H>T$, the magnetic susceptibility is a function of $H$, and it is more convenient to consider the magnetization in this case. Results of our analysis can be qualitatively summarized as follows.

In the case of the Dirac and Weyl semimetals, the magnetic susceptibility $\chi(\zeta)$ in the region of the weak magnetic fields exhibits the giant anomaly of the logarithmic type when the chemical potential $\zeta$ shifts with respect to the degeneracy energy $\varepsilon_d$, $\chi= A_D \ln(\varepsilon_0 /|\zeta -\varepsilon_d|)$. Here the constant $A_D$ is a combination of the   parameters characterizing the Dirac (Weyl) point, and the cut-off parameter  $\varepsilon_0$ is of the order of the energy spacing between $\varepsilon_d$ and other electron energy bands at this point of the Brillouin zone. In the weak magnetic fields, the magnitude of the anomaly is determined by the temperature, $\chi(\varepsilon_d) \approx A_D\ln(\varepsilon_0/T)$. In strong magnetic fields, when $\Delta \varepsilon_H$ becomes larger than $T$, the magnitude depends on $H$ logarithmically, $\chi(\varepsilon_d) \approx A_D\ln(\varepsilon_0 /\Delta\varepsilon_H)$, and hence the magnetization is proportional to $H\ln H$. When $\zeta$ is not close to the energy $\varepsilon_d$, in the strong magnetic fields the de Haas -van Alphen oscillations  appear. However, the phase of these oscillations is shifted by $\pi$ as compared to the usual case. It is also necessary to note that the giant anomaly in susceptibility is absent for type-II Weyl or Dirac semimetals.

In the case of the line node semimetals, the degeneracy energy $\varepsilon_d$ changes along the band-contact line in the interval $2\Delta\equiv \varepsilon_{max}- \varepsilon_{min}$ from its minimum value $\varepsilon_{min}$ to its maximum value $\varepsilon_{max}$. In the weak magnetic fields and at low temperatures,
the longitudinal magnetic susceptibility $\chi_{\parallel}(\zeta)$ exhibits the giant anomaly of the type $\chi_{\parallel}= A_{ln}/\sqrt{|\zeta -\varepsilon_c|}$ when $\zeta$ lies inside the interval $2\Delta$ and tends to one of the critical energies $\varepsilon_c$ ($\varepsilon_c=\varepsilon_{min}$ or   $\varepsilon_{max}$). Here $A_{ln}$ is a negative constant specified by certain parameters of the band-contact line. For $\zeta$ in the middle of the interval, one has $\chi_{\parallel} \sim  A_{ln}/\sqrt{\Delta}$, and $|\chi_{\parallel}|$ may be large if $\Delta$ is small. If $\zeta$ is outside the interval, the susceptibility is practically independent of the chemical potential. The divergence of $\chi_{\parallel}(\zeta)$ at $\zeta \to \varepsilon_c$ is cut off at $|\zeta-\varepsilon_c|\sim$  max($T,\Delta\varepsilon_H$). Hence, in the strong magnetic fields, when $T\ll \Delta\varepsilon_H\ll \Delta$, we arrive at $\chi_{\parallel}(\varepsilon_c)\approx A_{ln}/ \sqrt{\Delta\varepsilon_H} \propto H^{-1/4}$, and $M_{\parallel}\propto H^{3/4}$. When $H$ further increases so that $\Delta\varepsilon_H\gg T$ and $\Delta$, the magnetization is proportional to $H^{1/2}$. Interestingly, the angular dependences of the magnetization in the strong magnetic fields essentially differ from the appropriate dependences in the weak fields. As in the case of the Dirac semimetals, in the strong magnetic fields, the de Haas -van Alphen oscillations of the magnetization can appear for the line node semimetals, with the phase of the oscillations being shifted by $\pi$ as compared to the usual case.

Apart from the longitudinal magnetization $M_{\parallel}$, the magnetization components $M_{\perp}$ and $M_{\phi}$, which are perpendicular to ${\bf H}$, generally exist for the line node semimetals. These components generate the magnetic torques $M_{\perp}H$ and $M_{\phi}H$. The component $M_{\perp}$, which lies in the plane passing through ${\bf H}$ and the normal ${\bf z}$ to the plane of a closed band-contact line, is simply expressed in terms of $M_{\parallel}$, Eq.~(\ref{46}). The second perpendicular component $M_{\phi}$ appears only in strong magnetic fields, and it oscillates when ${\bf H}$ rotates about the normal ${\bf z}$ (${\bf z}$ usually coincides with a symmetry axis). The appearance of this $M_{\phi}$ in the strong magnetic fields is a distinguishing feature  of the band-contact lines.

Since the magnetization and magnetic susceptibility of the topological nodal semimetals are expressible in terms of the parameters characterizing the Dirac (Weyl) points and the band-contact lines, the magnetic measurements can be useful in investigating these semimetals.

\end{document}